%% file: main.tex
\documentclass[acmsmall,nonacm]{acmart}

\AtBeginDocument{%
  }

\begin{document}


\title{Usage of Virtual Reality in Combating Social Anxiety Disorders in Non-native English Speakers: A Survey}

\author{Siyi Zhang}
\email{szhan253@ucsc.edu}
\affiliation{%
  \institution{University of California, Santa Cruz}
  \city{Santa Cruz}
  \state{CA}
  \country{USA}
}

\author{Ayesha Khalid}
\affiliation{%
  \institution{University of California, Santa Cruz}
  \city{Santa Cruz}
  \state{CA}
  \country{USA}
}
\email{aykhalid@ucsc.edu}

\renewcommand{\shortauthors}{Zhang et al.}

\begin{abstract}
Social Anxiety Disorder (SAD) is a common yet underestimated mental health disorder. While non-native English speaker (NNES) students face public speaking, they are more likely to suffer some public speaking anxiety (PSA) due to linguistic and sociocultural differences \cite{cite1}. Virtual Reality (VR) technology has already benefitted social-emotional training. The core objective is to summarise the benefits and limitations of using VR technology to help NNES students practice and improve their public speaking skills. This is not a comprehensive survey of the literature. Instead, the selected papers are intended to reflect the current knowledge across various broad topics. Virtual Reality, Social Anxiety Disorder, Public Speaking Anxiety, English as a Second Language, and Non-native English speakers are the keywords used for searching mainly in the Academic Search Complete (ASC) database.
Compared with native English speaker (NES) students, NNES students have the potential to achieve better results when using VR technology for PSA social-emotional training.
\end{abstract}

\maketitle
\input{intro}
\input{background}
\input{literature}
\input{potential}
\input{challenges}
\input{conclusion}

\bibliographystyle{ACM-Reference-Format}
\bibliography{main}










\end{document}

%% file: intro.tex
\section{Introduction}
With the growing need for advanced language skills due to globalization and the growing interest in public speaking competitions, the importance of improving university students’ public speaking skills is increasing. The number of non-native English speakers (NNESs) enrolling in universities using English as their teaching language is also growing~\cite{cite2}. As public speaking is becoming more common in higher education and a mode of assessment as a part of final projects in many courses in different disciplines, higher education institutions have recognized the need for advanced speaking abilities for NNES international students. Social anxiety disorder (SAD), characterized by intense fear and avoidance of social or performance situations, significantly impacts individuals’ academic, professional, and personal lives~\cite{cite2, cite3}. This challenge is further compounded for students whose first language is not English as they navigate the dual hurdles of language proficiency and public speaking anxiety (PSA) ~\cite{cite1}.

Traditional methods of social anxiety treatment have typically relied on cognitive-behavioral therapies (CBT) and exposure-based interventions to help individuals confront and manage their fears~\cite{cite4}. While these approaches have shown some efficacy, there remains a need for innovative interventions that address the unique challenges faced by individuals with SAD.

In recent years, Virtual Reality (VR) has emerged as a promising tool for social-emotional skills training, offering a safe and controlled environment for individuals to practice social interactions and confront anxiety-provoking situations~\cite{cite5}. VR-based interventions have been shown to elicit physiological stress responses similar to real-world scenarios, providing a valuable platform for exposure therapy and skills development~\cite{cite6}.

The paper’s primary purpose is to highlight the potential advantages of using VR technology in social-emotional training to help NNES with SAD. The review begins with the basic SAD information and emphasizes the NNES group in the following discussion. The traditional psychology social training method, CBT, for individuals with SAD and how VR encompasses the benefits to help the social-emotional training based on the traditional method are also discussed in the paper.

%% file: background.tex
\section{Background}
\subsection{Social Anxiety Disorder}
C. Steinert et al. reviewed several SAD-related papers and introduced to us what we know today about SAD. SAD, also known as social phobia, is distinguished as the most frequent anxiety disorder and ranks as the second most prevalent among all DSM-IV disorders. It is characterized by an intense fear triggered by social situations where the individual fears scrutiny by others. This fear can manifest in various social interactions, such as communicating, eating, or speaking in public, and can lead to avoidance behaviors or substantial discomfort when avoidance isn’t feasible. The avoidance linked with SAD can significantly impair social and occupational functioning, potentially leading to social isolation, diminished quality of life, increased unemployment, and reliance on social welfare. Moreover, SAD often co-occurs with other mental disorders, including depression, other anxiety disorders, substance abuse, and personality disorders, with comorbidity rates ranging from approximately 69\% to 99\%. The high prevalence of comorbid conditions further complicates the distinction between outcomes attributable to SAD and those related to accompanying disorders. Notably, only about 5\% of individuals with SAD seek appropriate help, partly because they might not perceive themselves as ill but rather as excessively shy. It’s often the emergence of comorbid disorders, which intensify psychological distress, that prompts those affected by SAD to seek medical assistance~\cite{cite7}.

\subsection{Public Speaking Disorder}
Social anxiety disorder (SAD) represents a significant and underestimated mental health challenge with a lifetime prevalence of 12.1\% and is characterized by intense fear and discomfort in social interactions. Research suggests that individuals with SAD hold negative views about their personality, social skills, and fear of being embarrassed during social interactions. Public speaking anxiety (PSA) is one of the most common fears in teenagers and adults~\cite{cite8}. PSA is classified as a type of performance anxiety. Individuals with PSA overestimate the expectations of being negatively evaluated when speaking in front of others and underestimate their capability. This fear significantly affects individuals, leading to a wide array of negative outcomes, including social isolation, impaired functioning, and substantial distress~\cite{cite9}. The underlying mechanism of this anxiety involves individuals’ persistent concern over being negatively judged, leading them to anticipate adverse evaluations in social situations~\cite{cite10}. Such anticipatory anxiety not only heightens their distress during social interactions but also discourages them from engaging in social activities, potentially resulting in social withdrawal~\cite{cite11}.

The fear of negative evaluation is central to understanding public speaking anxiety. Individuals with this anxiety often hold negative self-views regarding their appearance, public speaking abilities, personality, and social skills~\cite{cite12}. These negative self-conceptions heighten their sensitivity to potential criticism and rejection, fueling their fear of being evaluated unfavorably. The expectation of negative evaluation, whether accurate or exaggerated, triggers a cycle of anxiety that can be debilitating~\cite{cite13}.

Public speaking anxiety has significant implications for individuals’ social and professional lives. It can hinder academic and career progress, as affected individuals may avoid situations where speaking in front of others is required, leading to missed opportunities for advancement~\cite{cite14}. Furthermore, the stress associated with this anxiety can contribute to a range of psychiatric complications, including depression, substance abuse, and even suicidal ideation, as chronic stress and isolation exacerbate underlying mental health issues~\cite{cite15}.

Research indicates that CBT, including exposure to feared situations, cognitive restructuring of maladaptive thoughts, and the development of coping strategies, can be effective in addressing public speaking anxiety~\cite{cite16}. By targeting the fear of negative evaluation and helping individuals to develop more positive self-conceptions and realistic assessments of social threats, these interventions can significantly reduce the distress associated with public speaking and improve overall functioning~\cite{cite17}.

\subsection{Challenges of Public Speaking for NNES Students}
NNES students face linguistic, sociocultural, and psychological challenges in preparing for and delivering academic presentations. These students often spend significantly more time than NES preparing for academic presentations due to language limitations and unfamiliarity with the sociocultural norms of academic presentations in English-speaking contexts. Common challenges include the lack of experience and skills for extemporaneous speaking, differences in classroom interaction patterns, and the struggle to sound ’smart’ and articulate complex ideas clearly~\cite{cite1}.

Additionally, the fear of making linguistic mistakes and not being able to express themselves clearly can lead to heightened self-consciousness and avoidance of social interactions~\cite{cite18}. This fear is closely tied to the fear of negative evaluation, a central component of SAD, as these students may worry excessively about being judged for their language proficiency~\cite{cite8}.

Cultural differences further exacerbate these challenges. NNES students must navigate the complex task of adjusting to a new cultural environment, which can be daunting. Cultural norms and communication styles vary widely, and misunderstandings or misinterpretations can contribute to feelings of alienation and inadequacy~\cite{cite19}. These experiences can reinforce negative self- perceptions and the fear of negative evaluation, fueling social anxiety.

The academic environment itself poses additional pressures. NNES students often experience anxiety related to academic performance, fearing that their language skills may negatively impact their grades and future opportunities~\cite{cite20}. This academic anxiety can compound social anxiety, as students may avoid participating in class discussions or group projects, further isolating themselves.

Moreover, the lack of social support in the new environment can make NNES students more susceptible to SAD. Social support is a critical buffer against stress and anxiety~\cite{cite21}. Without a strong support network, NNES students may struggle to cope with the challenges they face, increasing their risk of developing SAD.

\subsection{Traditional Methods}
Research suggests medication and psychotherapy are effective treatments for individuals with SAD. CBT is the consensus in the field of psychotherapy and addresses negative patterns of thought and behavior to effect positive change through core components like cognitive restructuring, exposure therapy, and social skills training. CBT is based on the cognitive model of emotional response, which posits that our thoughts, feelings, and behaviors are interconnected and that changing negative thought patterns and behaviors can lead to changes in our feelings~\cite{cite22}.

One of the critical components of CBT is cognitive restructuring, which involves identifying irrational or maladaptive thoughts and challenging their accuracy. This process helps individuals develop more balanced and realistic thoughts, leading to healthier emotional responses. Behavioral interventions in CBT may include exposure therapy, where individuals gradually face their fears in a controlled manner to diminish their anxiety response, and behavioral activation, particularly effective in treating depression, which involves engaging in activities that are pleasurable or provide a sense of achievement to counteract the inertia that often accompanies depression~\cite{cite23}.

CBT has been adapted for individual, group, and self-help formats, including online platforms, making it accessible to a wide range of individuals. Its efficacy has been supported by extensive research, making it the gold standard treatment for numerous psychological disorders. Importantly, CBT empowers individuals with the skills to manage their symptoms and prevent relapse, focusing on making long-term changes rather than merely providing temporary relief~\cite{cite24}.

The effectiveness of fear of negative evaluation (FNE) and self-verification process for individuals with SAD are also examined. The findings revealed that self-compassion training could lead to significant improvements in SAD symptoms, pointing to the therapeutic value of cultivating a compassionate self-relationship. This research suggests that fostering self-compassion could be a promising strategy for individuals struggling with public speaking anxiety, offering a pathway to reduce anxiety symptoms by promoting a kinder, non-judgmental attitude towards oneself~\cite{cite25}.

\subsection{The Role of Fear of Negative Evaluation (FNE)}
FNE, a core aspect of SAD, plays a significant role in the distress experienced by individuals during public speaking. NNES students with SAD may overestimate the potential for negative evaluation while underestimating their own performance, leading to heightened anxiety and avoidance behaviors. Addressing FNE through therapeutic interventions is crucial for reducing social anxiety and improving public speaking performance~\cite{cite25}.

\subsection{Virtual Reality(VR) and Virtual Environments(VE) in Social Training}
VR creates a three-dimensional computer-based virtual environment (VE) immersive world to let the users move through and interact with the objects in the world. Research has already suggested using VR in the fields of education and rehabilitation, such as helping to do social training for individuals with autism spectrum disorder, reducing vertigo and fly phobia, helping children with learning disabilities to develop everyday skills, and helping children with visual impairments~\cite{cite5}.

The immersive nature of VR technology helps to simulate a real-world-like environment but without the real potential harm that the users might suffer in the real world. The digital environment also gives the researchers the ability to control exposure to the social skills that the user needs to practice, which provides the ability to customize scenarios based on different users. That is also a way of letting the users engage in social-emotional training without real-world consequences. Compared to the real-world training method, the environment is easier to repeat using VR technology. Although VR technology is a new social-emotional way compared with traditional psychology methods, it still complements traditional therapies, such as CBT. CBT is still mentioned in most VR social-emotional training studies.

Recent advances in VRE also have shown promise in treating SAD, especially fears associated with public speaking. VRE allows for controlled, repeatable, and customizable exposure to public speaking scenarios, providing a practical and cost-effective treatment option. Studies have demonstrated that VRE can effectively reduce Public speaking anxiety by allowing systematic exposure and habituation to the feared stimuli~\cite{cite5}.

%% file: literature.tex
\section{Existing Research on Using VR for SAD}
\subsection{SAD In Psychology}
The prevailing perspective in psychology on SAD covers a wide array of research areas. CBT has been proven effective and is widely accepted as a method for ameliorating SAD. Studies also delve into topics such as self-verification processes, socialization through academic discourse, the employment of metaphors to alleviate public speaking anxiety and the impact of perfectionism. Insights from these psychological viewpoints can be instrumental in informing the development of interactive VE.

Howarth and Forbes explored the role of self- verification processes in social anxiety, finding that individuals with higher social anxiety levels were more comfortable with negative feedback and viewed it as more accurate than positive feedback. Their study suggests that self-verification processes, which involve seeking feedback that confirms one’s self-view, operate in social anxiety, contributing to the maintenance of negative self-conceptions and increased social anxiety~\cite{cite25}.

Kahlon et al. conducted a study on the effect of perfectionism on interventions for public speaking anxiety in adolescents. The findings indicated that interventions did not lead to a reduction in perfectionism, but a decrease in perfectionism during follow-up was associated with a more significant reduction in social anxiety symptoms. High pre-treatment levels of perfectionism were linked to poorer outcomes, suggesting the need to assess and address perfectionism in interventions~\cite{cite2}.

Sapach and Carleton explored the effectiveness of self-compassion training for individuals with Social Anxiety Disorder, finding that self-compassion training was statistically superior in improving outcomes compared to a waitlist control condition. Their study provided preliminary evidence for the effectiveness of self-help self-compassion training programs in managing clinically significant SAD symptoms~\cite{cite4}.

These studies collectively illustrate the complexity of addressing social anxiety disorder through various approaches, including the exploration of self-verification processes, academic socialization challenges, the use of metaphors to understand public speaking anxiety, the impact of perfectionism, and the efficacy of school-based and self-compassion interventions. All the approaches could help understand SAD better and be used while designing VE interactions.

\subsection{VR Usage for SAD}
\begin{figure}[ht]
  \centering
  \includegraphics[width=0.5\linewidth]{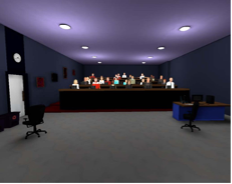}
  \caption{This VE simulates the participant view of an auditorium-style room created by R. Sigurvinsdottir et al. and used in their research.}
  \Description{This VE simulates the participant view of an auditorium-style room created by R. Sigurvinsdottir et al. and used in their research.}
  \label{fig:ve_sim}
\end{figure}

In most research using VR technology in their social-emotional training for individuals with SAD, the VE is set up to simulate a public speaking scenario where the player stands on a stage facing the audience. The setting is designed to replicate aspects of public speaking engagements to simulate a real-world environment to support the users’ practice. The environment normally includes some realistic 3D models of podiums, stages, and audience seating. Some researchers also add audience sound effects to make the VE more immersive.

The study done by R. Sigurvinsdottir et al. investigated the impact of VR on inducing distress among individuals with SAD, focusing on the FNE as a mediator between social anxiety and distress. The VR environment, simulating a classroom scenario for public speaking, effectively induced distress among participants. The findings suggest FNE could be a critical treatment target within VR or in-person to reduce distress related to audience presentation. The use of VE by R. Sigurvinsdottir et al. in the study is also described in detail in the paper. They built a part of the school’s VE that comprised the hallway, classroom, and connecting door, as shown in Figure \ref{fig:ve_sim}. Common surrounding background sounds effective, such as some paper-turning sounds that are commonly heard in a classroom, also added in the VE to help make the environment more immersive~\cite{cite12}.

M. Takac et al. focused on distress habituation across repeated VR training sessions for public speaking. Multiple scenarios are included in the study to increase the diversity of testing environments. Despite the absence of within-scenario habituation, between-scenario habituation was observed, indicating the importance of multiple exposure sessions. Long-term follow-ups are included in the study twice. The first one is after three months, and the second follow-up is in the sixth year, which strongly supports the study’s highlights that VR’s efficacy in eliciting public speaking distress and facilitating habituation, suggesting brief, repeated VR exposures can significantly impact managing public speaking anxiety~\cite{cite26}.

M. P. A. Howarth and M. Forbes use randomized controlled trials to evaluate the efficacy of VR exposure embedded in CBT against traditional in vivo exposure for treating SAD. Self-administered assessments were conducted for each participant before and after each treatment. There is also a long-term follow-up to keep track of the effect of the CBT therapy. The study found improvements across all measures for both CBT groups compared to a waiting list control, with VR exposure demonstrating significant practicality for therapists over in vivo exposure. The findings advocate for VR’s advantages as a treatment medium, potentially offering solutions for treatment avoidance and enhancing therapy efficiency and practicality~\cite{cite27}.

Some other factors are discussed in the current existing studies. R. Sigurvinsdottir et al. mention that although they added some common classroom noise in the VE, no participant reported paying attention to those sounds during the test. Additionally, R. Sigurvinsdottir et al. also consider waiting time in the classroom before the participants do their presentation as a factor and try to analyze whether the waiting time would affect the anxiety for public speaking, but there is not much evidence shown in the examined variables could tell the waiting time is an effective factor ~\cite{cite12}.

%% file: potential.tex
\section{The Potential for VR with NNES Students}
The exploration of VR technology in the treatment and training for SAD and PSA reveals its potential to offer a controlled, immersive environment where individuals can confront and habituate to anxiety-provoking scenarios. The capacity of VR to simulate public speaking settings provides a safe space for NNES students to practice and refine their speaking skills without the immediate pressure of real-life social evaluations~\cite{cite12,cite26}. This aspect is particularly crucial for NNES students who face additional linguistic and cultural barriers that can exacerbate their anxiety and hinder their participation in conventional training settings.

NNES students grapple with the dual challenge of mastering a second language while also navigating the complexities of public speaking. The integration of VR into their training regimen can be particularly beneficial, as it allows for repeated exposure to speaking situations in a controlled environment, thus helping to build confidence and reduce linguistic anxiety~\cite{cite1,cite3}. Furthermore, VR can be programmed to introduce cultural nuances and social cues relevant to various English-speaking contexts, aiding students in better understanding and adapting to different social norms and expectations.

Comparing VR-based interventions to traditional methods such as Cognitive-Behavioral Therapy (CBT) highlights several potential advantages. While CBT remains the gold standard for treating SAD, incorporating VR could enhance the efficacy of exposure therapy by providing more versatile and customizable exposure scenarios. This adaptability is particularly beneficial for NNES students, who may require specific types of exposure that are difficult to replicate in real life~\cite{cite4,cite27}. Moreover, VR’s capacity to simulate a wide range of social and cultural contexts can offer more comprehensive training, addressing both the fear of negative evaluation (FNE) and the broader spectrum of social anxieties faced by these students.

The promising results from VR-based studies indicate a need for further research to explore the long-term efficacy of VR interventions and their potential integration with traditional therapies. As VR technology continues to evolve, future studies should also examine the impact of more immersive and interactive VR experiences on reducing PSA and SAD symptoms among NNES students. Additionally, considering the role of self compassion and perfectionism in mediating social anxiety symptoms, future interventions could benefit from incorporating elements that foster self-compassion within the VR training modules~\cite{cite2,cite25}.

%% file: challenges.tex
\section{Challenges of Using VR in Social-Emotional Training}
Although the benefit of using VR technology in social-emotional training, especially for NNES students, there is still some shortage of using VR technology. Despite the potential benefits, the initial cost of VR equipment can be a barrier for some learners and institutions. Not every NNES student or individual with SAD has the resources to purchase VR equipment for personal practice, which could limit the widespread adoption of VRbased training solutions. Research indicates that repeated exposure to VR scenarios is necessary for reducing public speaking anxiety effectively~\cite{cite26}. This need for habituation highlights the importance of sustained and repeated use of VR training programs, which might require significant time and commitment from the users.

Some users experience cybersickness, which includes symptoms like headaches and dizziness while using VR equipment. This issue can significantly hinder the training process, especially for individuals with SAD, who might already be prone to anxiety-related physical symptoms~\cite{cite12}. The effectiveness of VR technology in SAD training is constrained by the prevalence of cybersickness among users. Addressing this challenge requires ongoing research and development to create more comfortable and user-friendly VR experiences. With the development of technology, there is some more comfortable equipment that could provide VE been produced. Such as Apple Vision Pro reduces dizziness while using related equipment is mentioned in most online user experience feedback. Although it is even more expensive compared with other VR equipment, it still shows the potential of the development of technology that could help us to create a better tool to support social-emotional training.

%% file: conclusion.tex
\section{Conclusion}
In conclusion, although there is still room to improve in using VR technology for social-emotional training, especially with NNES students who have SAD, it offers a novel and promising avenue for social-emotional training, particularly for NNES students facing SAD and PSA. By providing a safe, controlled environment for repeated practice and exposure, VR has the potential to significantly mitigate the anxieties associated with public speaking and social interactions in a second language context. As this field continues to develop, further research will be essential in optimizing VR interventions for maximum benefit to individuals with SAD, thereby enhancing their academic, professional, and personal lives.